\documentclass[a4paper,10pt]{IEEEtran}

\usepackage{balance}
\usepackage{graphics}
\usepackage{graphicx}
\usepackage[utf8]{inputenc}
\usepackage{fancyvrb}
\usepackage{listings}
\usepackage{color}
\usepackage{paralist}
\usepackage{xspace}
\usepackage{url}
\usepackage[bookmarks=false]{hyperref}
\usepackage[T1]{fontenc}
\usepackage{booktabs}
\usepackage{cite}
\usepackage{amsmath}
\usepackage{amssymb}

\newcommand\f{\mathit{F}}

\newcommand\fm{\mathit{FM}}
\newcommand\maxp{\mathbb{P}}
\newcommand\impl[1]{\mathit{impl}(#1)}
\newcommand\spec[1]{\mathit{spec}(#1)}
\newcommand\varenc[2]{\mathit{var\_enc}(#1,#2)}
\newcommand\sel[2]{\mathit{select}(#1,#2)}
\newcommand\refines[3]{\mathit{refines}(#1,#2,#3)}

\newcommand\fheader[1]{\hfill{\scriptsize Feature \em #1~~~\vskip -1.5ex}}

\newcommand\fh{\textsc{FeatureHouse}\xspace}

\newcommand\fc{\textsc{SPLverifier}\xspace}
\newcommand\code[1]{{\small\textsf{#1}}}
\newcommand\feature[1]{\emph{#1}}

\newcommand{\smallsec}[1]{\paragraph*{#1}}

\lstdefinelanguage{ar}[ANSI]{c++}%
  {morekeywords={introduction,shadow,automaton,fail,refines,super,Super,original,this,layer,pointcut,call,aspect,execution,advice,around,before,after,execution,this,target,within,args,declare,parents,implements,throws,returning,throwing,throw,synchronized,boolean,proceed,cflow,cflowbelow,null,abstract,extends,interface,instanceof,privileged,cclass,document,grammar,overrides},%
   sensitive=f
   }[keywords,comments,strings]%

\begin{document}

\pagestyle{empty}
\begin{minipage}{17cm}
\begin{center}
~\\[1cm]
\Huge{Feature-Aware Verification}
\\[2cm]
\large{Sven Apel\,$^1$, Hendrik Speidel\,$^1$, Philipp Wendler\,$^1$, \\
         Alexander von Rhein\,$^1$, and Dirk Beyer\,$^{1,2}$}
\\[1cm]
\normalsize
{$^1$\,University of Passau, Germany}\\
{$^2$\,Simon Fraser University, B.C., Canada}\\[5cm]

\hspace{-5mm}
\includegraphics[scale=0.2]{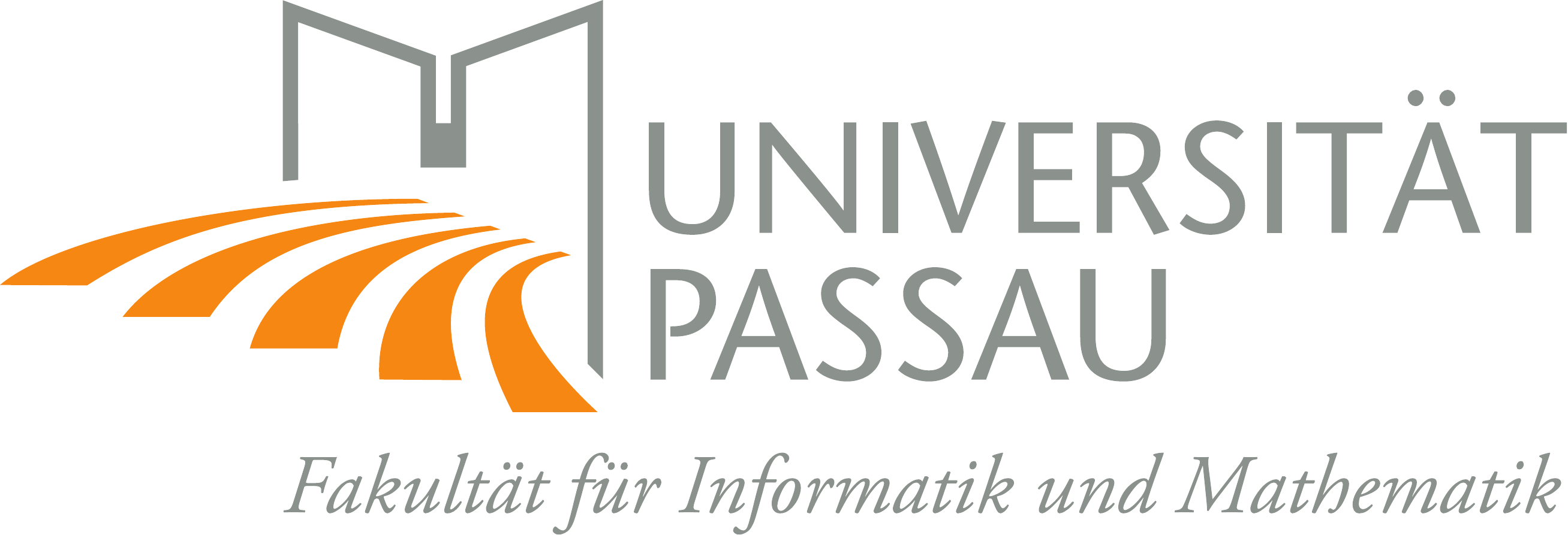} \\[1cm]
Technical Report, Number MIP-1105\\
Department of Computer Science and Mathematics\\
University of Passau, Germany\\
September 2011
\end{center}
\end{minipage}

\title{Feature-Aware Verification
  \thanks{An abbreviated version of this article appeared in Proc.~ASE 2011~\cite{ASE11}.}
}

\author{{Sven Apel\,$^1$, Hendrik Speidel\,$^1$, Philipp Wendler\,$^1$, 
         Alexander von Rhein\,$^1$, and Dirk Beyer\,$^{1,2}$}
\vspace{3mm}\\
{$^1$\,University of Passau, Germany}\\
{$^2$\,Simon Fraser University, B.C., Canada}\\
}

\maketitle
\thispagestyle{empty}
\setcounter{page}{1}
\pagestyle{plain}

\begin{abstract}
% Introducing the topic and stating the problem.
A software product line is a set of software products that are distinguished in terms of features 
(i.e., end-user--visible units of behavior). 
Feature interactions ---situations in which the combination of features leads to emergent 
and possibly critical behavior---
are a major source of failures in software product lines.
% Explaining the contributions.
We explore how \emph{feature-aware verification} can improve the automatic detection of 
feature interactions in software product lines. 
Feature-aware verification uses product-line verification techniques and supports the specification of feature properties 
along with the features in separate and composable units. 
It integrates the technique of \emph{variability encoding} to verify a product line 
without generating and checking a possibly exponential number of feature combinations.
% Implementing and experimenting.
We developed the tool suite \fc for feature-aware verification, which is based on standard 
model-checking technology.
We applied it to an e-mail system that incorporates domain knowledge of AT\&T.
We found that feature interactions can be detected automatically based on specifications that 
have only feature-local knowledge, and that variability encoding 
significantly improves the verification performance
when proving the absence of interactions.

\end{abstract}

\section{Introduction}
A \emph{software product line} is a family of software products that share a common set of features and differ in others~\cite{CN01}.
A \emph{feature} is an end-user--visible behavior of a software product that is of interest for some stakeholder.
A \emph{feature interaction} is a situation in which the composition of several features leads to emergent behavior that does not occur when one of them is absent.
The feature-interaction problem (i.e., the problem of predicting and detecting feature interactions) has been studied and addressed before and is still a major challenge~\cite{CKM+03}.

A \emph{feature-oriented} product line is composed of feature modules that encapsulate the code of each feature into a separate and composable unit~\cite{AK09fosd}. 
While this approach increases variability and reusability~\cite{AK09fosd}, it makes the detection of feature interactions difficult, because typically many feature combinations are possible, and an interaction may occur only in some of them.

Our aim is to explore how product-line--verification techniques~\cite{LTP09,CHS+10,CHS+11} (i.e., efficiently verifying that all products of a product line satisfy their specification) can be used to automatically detect feature interactions.
Especially, we concentrate on two challenges that arise in feature-oriented software product lines: 
A first challenge, which was formulated by Hall~\cite{Hal05}, is how to detect feature interactions based on specifications that do not have global system knowledge.
The background is that the specification of a feature should not need to be aware of all other features of the system. 
It is desirable to specify \emph{and} implement features in separate and composable units, while still being able to detect feature interactions, which typically emerge when multiple features are combined~\cite{Hal05,ASL+10}.

A second challenge, which applies to product-line analysis in general~\cite{PS08,LTP09,CHS+10,AKG+10,CHS+11,KAT+11}, is to detect feature interactions without the need of generating and checking all individual products. 
Typically, many different feature combinations are possible, so detecting feature interactions by generating all possible combinations may not be feasible.

We call the approach of verifying the absence (or detecting the presence) of feature interactions in feature-oriented product lines \emph{feature-aware verification}. 
We base it on a number of ingredients.
First, we provide a specification language to specify a feature's temporal properties in a separate and composable unit (along with its implementation).
Second, we use the technique of \emph{variability encoding} (which is based on configuration lifting~\cite{PS08}) to verify a complete product line in a single run ensuring that all possible feature combinations are free of critical feature interactions.
Third, we use off-the-shelf model-checking techniques, rather than relying on modifications and extensions of existing model checkers. This has the benefit that we can experiment with different model checkers and profit from recent developments in this field.

We have developed the tool chain \fc for feature-aware verification, 
and we use it in a case study ---an e-mail client that was developed as a product line--- 
to investigate the potential of feature-aware verification for detecting feature interactions. 
We base our study on Hall's e-mail system specification, which is a test-bed for feature interactions and incorporates domain knowledge of AT\&T:
it contains several realistic and unintuitive feature interactions and it has been used by other researchers in this area~\cite{LKF05}.

In summary, we contribute the following:%
\footnote{All data and results are available on the Web: \url{http://fosd.net/FAV/}.}
\begin{compactitem}
\item We present feature-aware verification, an approach to detect feature interactions using composable feature implementations and specifications.
It uses product-line verification techniques, variability encoding, and off-the-shelf model checking technology.
\item We formalize key aspects of feature-aware verification including variability encoding, and we provide a proof of its correctness, 
which has not been provided by previous work~\cite{PS08}.
\item By means of a case study that incorporates the domain knowledge of AT\&T on e-mail systems, we explore how feature-aware verification is able to automatically detect feature interactions in feature-oriented product lines.
We found that feature-aware verification is able to detect all interactions in Hall's e-mail client based on feature specifications that have only local knowledge.
\item Based on a number of experiments, we derive a model that describes when variability encoding is beneficial. 
We found that variability encoding improves the verification performance if the task is to prove the absence of unsafe feature interactions or if the product line 
contains only few interactions that violate the specification.
\end{compactitem}
In previous work, a modeling language and a model analyzer were used to detect unsafe feature interactions in feature-oriented design~\cite{ASL+10}, however, without considering the challenges of product lines and without a formal model and measurements. A short version of this report has been published in the ASE'11 proceedings~\cite{ASE11}.

\section{Background and Problem Statement}
\label{sec:background}
The goal of feature orientation is to make features explicit in design and code, for example, in the form of composable feature modules~\cite{AK09fosd}.
Products are composed from feature modules by superimposition~\cite{AKL09fh},
which is a language-independent form of deep mixin composition~\cite{Hut09}.
Basically, superimposition merges the code of all features recursively based on nominal and structural similarity.
Typically, there is a total order of feature modules defined, because feature composition is not generally commutative~\cite{ALM+10scp}.
In our case study, we use the tool \fh~\cite{AKL09fh} for composition.

In Figure~\ref{fig:e-mail_client_example}, we depict excerpts of four feature modules 
taken from our case study.
Feature \feature{EmailClient} implements a basic e-mail client, feature \feature{Encrypt} 
encrypts outgoing \mbox{e-mails}, feature \feature{Decrypt} decrypts incoming e-mails, and feature \feature{Forward} forwards incoming e-mails to another host. Note that encryption and decryption are asymmetric and rely on the availability of proper keys --- a circumstance that gives rise to a feature interaction, as we will explain shortly.

\begin{figure}[h!]
\centering
\fheader{EMailClient}
\begin{lstlisting}[firstnumber=1,frame=top|bottom,basicstyle=\sf\scriptsize,
keywordstyle=\sf\scriptsize\bfseries,commentstyle=\sf\scriptsize\it]{ar}
// representation of e-mail
struct email {
  int id; char *from; char *to; char *subject; char *body;
};

// outgoing e-mails are processed by this function before they leave the system
void outgoing (struct client *client, struct email *msg) { ... }

// incoming e-mails reach the client at this point and are stored in a mailbox
void incoming (struct client *client, struct email *msg) { ... }
\end{lstlisting}
\vspace{1ex}
\fheader{Encrypt}
\begin{lstlisting}[frame=top|bottom,basicstyle=\sf\scriptsize,
keywordstyle=\sf\scriptsize\bfseries,commentstyle=\sf\scriptsize\it]{ar}
// extending the e-mail structure by information on encryption
struct email {
  int isEncrypted;
  char *encryptionKey;
};

// encrypt a given e-mail, if the public key of the receiver is known
void encrypt (struct client *client, struct email *msg) { ... }

// override function outgoing to encrypt e-mails before they are sent
void outgoing (struct client *client, struct email *msg) {
  encrypt (client, msg);
  original (client, msg); // invoke the overridden function
}
\end{lstlisting}
\vspace{1ex}
\fheader{Decrypt}
\begin{lstlisting}[frame=top|bottom,basicstyle=\sf\scriptsize,
keywordstyle=\sf\scriptsize\bfseries,commentstyle=\sf\scriptsize\it]{ar}
// decrypt a given e-mail
void decrypt (struct client *client, struct email *msg) { ... }

// override function incoming to decrypt encrypted incoming e-mails 
void incoming (struct client *client, struct email *msg) {
  decrypt (client, msg);
  original (client, msg); // invoke the overridden function
}
\end{lstlisting}
\vspace{1ex}
\fheader{Forward}
\begin{lstlisting}[frame=top|bottom,basicstyle=\sf\scriptsize,
keywordstyle=\sf\scriptsize\bfseries,commentstyle=\sf\scriptsize\it]{ar}
// forward an e-mail to another host
void forward (struct client *client, struct email *msg) { ... }

// override function incoming to forward e-mails automatically
void incoming (struct client *client, struct email *msg) {
  forward (client, msg);
  original (client, msg); // invoke the overridden function
}
\end{lstlisting}
\caption{A feature-oriented implementation of an e-mail client in C (excerpt).}
\label{fig:e-mail_client_example}
\end{figure}

\feature{EMailClient} is the base feature in our example.
It introduces a structure \code{email} for representing e-mails and the two functions \code{outgoing} and \code{incoming} for handling incoming and outgoing e-mails.
Composing it with feature \feature{Encrypt}, the existing structure \code{email} is extended by the 
two new fields \code{isEncrypted} and \code{encryptionKey}, function \code{encrypt} is added, 
and the existing function \code{outgoing} is overridden to intercept outgoing e-mails and to encrypt them using function \code{encrypt};
the keyword \code{original} is used to invoke the overridden function. 
Similarly, feature \feature{Decrypt} introduces a function \code{decrypt} and overrides the existing function \code{incoming} to intercept and decrypt incoming e-mails.
Finally, feature~\feature{Forward} introduces a function \code{forward} and overrides the existing function \code{incoming} to forward incoming e-mails to another host.

As there are different feature-oriented languages and tools available~\cite{ALM+10scp}, we concentrate on a common set of functionality: a feature module may add new fields, functions, and structures as well as refine existing functions by overriding.

Typically, products can be composed from features in different combinations. 
The compositional flexibility gives rise to feature interactions. 
A feature interaction is a situation in which new behavior emerges from the composition of two or more features that cannot easily be deduced from the behavior of the features involved. 
The emergent behavior can be undesired and associated with unexpected program states~\cite{CKM+03}.

While the features \feature{Encrypt} and \feature{Decrypt} of the e-mail example depend on each other (they share common data structures and functions) and should only be selected together, feature \feature{Forward} has been developed independently of the two, only based on feature \feature{EMailClient}.
The composition of all four features leads to an undesired feature interaction.%
The interaction occurs if one host sends an encrypted e-mail to a second host that forwards the e-mail automatically to a third host.
If the second host does not have the public key of the third host, it forwards the e-mail in plain text (\feature{Forward} does not know whether an e-mail is encrypted).
This situation violates the specification of feature \feature{Encrypt}, which states that e-mails that have been encrypted initially must never be sent unencrypted over the network.%
\footnote{Of course, we could implement or specify the features differently to fix the interaction, but we base the example on the work of Hall~\cite{Hal05}, which captures the essence of real-world feature interactions in e-mail systems.}

Hall notes that the detection of feature interactions based on \emph{feature-local} specifications is an open problem~\cite{Hal05}.
That is, the specification of a feature should not necessarily be aware of all other features of the system, but only of the ones it uses and extends directly.
In our example, we need a specification of the desired behavior of feature \feature{Encrypt} that states that e-mails that are received in encrypted form must not be sent in plain text --- without referring to other independently developed features such as \feature{Forward}.

Note that, even if there is a feature model that describes the domain dependencies between features~\cite{CE00}, it typically does not cover implementation-level dependencies that may lead to inadvertent feature interactions at runtime~\cite{TBK+07}.
Furthermore, in a scenario with distributed feature composition there is no global feature model --- features are developed in isolation and composed in an ad-hoc manner~\cite{JZ98}.
Hence, we need a mechanism that is able to detect feature interactions automatically based on the specifications and implementations of the features involved.

\section{Specifying Features}
\label{sec:specifying}
To be able to reason about feature interactions, each feature needs a formal specification of its behavior and the constraints that have to be fulfilled if it is selected (i.e., if it is present in the generated product).
A key goal of feature-oriented programming is to implement and specify features in separate and composable units.
Ideally, a feature's specification refers only to itself and a certain basis (i.e., the features that it extends and uses directly).
We would like to explore to what extent this is possible. 
Beside missing global domain knowledge, scalability is a motivation for feature-local specifications.
A~system in which every feature has to be aware of every other feature does not scale well with regard to program comprehension when the number of features increases. 

\begin{figure}[tbh]
\centering
\begin{lstlisting}[firstnumber=1,frame=top|bottom,basicstyle=\sf\scriptsize,
keywordstyle=\sf\scriptsize\bfseries,commentstyle=\sf\scriptsize\it]{ada}
// automaton definitions
auto_decl : "$\mbox{\bfseries automaton}$" auto_name "{" intro_decl? intercept_decl+ "}";

// introductions
intro_decl : "$\mbox{\bfseries introduction}$" "{" shadow_decl* c_decl* "}";

// shadow declarations
shadow_decl : "$\mbox{\bfseries shadow}$" c_struct_decl;

// function-execution interceptions
intercept_decl : ("$\mbox{\bfseries before}$" event_decl) | ("$\mbox{\bfseries after}$" (return_name "=")? event_decl);

// event declarations
event_decl : type event_name "(" param_list ")" c_func_body;
\end{lstlisting}
\caption{Grammar of automata-based specifications of features (simplified).}
\label{fig:grammar}
\end{figure}

We have developed a language to specify features in separate and composable units.
We define its syntax in Figure~\ref{fig:grammar}. 
A feature is specified by one or more automata, declared by keyword \code{automaton}.
Keyword \code{introduction} can be used to introduce auxiliary functions and 
structures (rule \code{c\_decl} refers to C declarations), and 
keyword \code{shadow} adds members to existing C structures (\code{c\_struct\_decl} 
refers to C structure declarations) that are visible \emph{only} to the automaton.
Both constructs are used to make the automata stateful.
The keywords \code{before} and \code{after} define events to intercept function executions.
In the body of the event declaration (\code{c\_func\_body} refers to C function bodies), 
the developer can use the keyword \code{fail} (not shown in Fig.~\ref{fig:grammar}) to indicate that the system reaches an error state, 
which we use to indicate a feature interaction.

An automaton specifies a safety property of the behavior of a feature.
That is, it defines in which circumstances related to the feature the execution of the overall system reaches an error state (fail); 
all other behaviors are accepted and thus considered safe.%
\footnote{To avoid situations in which an automaton modifies a program in an undesired way or even interferes with other automata, we require automata to be free of side effects.}
The automata language of Figure~\ref{fig:grammar} represents the subset of 
linear temporal logic that is concerned with safety properties.

Note that, for now, we do not allow a specification of one feature to extend or modify the specification of another feature. 
While we can imagine some examples for which this might be useful, we stick with the conservative approach to avoid unintended interactions at the level of specifications (there is the danger to lift the feature-interaction problem from the implementation level to the specification level).

\smallsec{Example}
In Figure~\ref{fig:encrypt_spec}, we show the specification of feature~\feature{Encrypt}.
When the client receives an encrypted e-mail (lines~6--8), the status (encrypted or not) of the e-mail is stored 
(line~7) into a field that has been attached as a shadow to structure \code{email} (line~3).
When an e-mail that was encrypted leaves the system (lines~10--12), it must still be encrypted; 
if not, the e-mail client reaches an error state defined by keyword \code{fail} (line~11).
This specification is based on the work of Hall~\cite{Hal05}.
It is local in the sense that it does not know anything about the changes feature 
\feature{Forward} makes to the e-mail system.
But still, it can be used to detect the interaction between
\feature{Encrypt} and \feature{Forward}.
\begin{figure}[tbh]
\centering
\begin{lstlisting}[firstnumber=1,frame=top|bottom,basicstyle=\sf\scriptsize,
keywordstyle=\sf\scriptsize\bfseries,commentstyle=\sf\scriptsize\it]{ar}
automaton EncryptSpec {
  introduction {
    shadow struct email { int in_encrypted; };
  }

  before void incoming(_:struct client*, msg:struct email*) {
    msg->in_encrypted = isEncrypted(msg);
  }

  after void outgoing(_:struct client*, msg:struct email*) {
    if(msg->in_encrypted != 0 && !isEncrypted(msg)) { fail; }
  }
}
\end{lstlisting}
\caption{Automaton-based specification of feature~\feature{Encrypt}.}
\label{fig:encrypt_spec}
\end{figure}

\section{Detecting Interactions}

Based on feature-local specifications, there are two options of detecting feature interactions 
in a product line: (1) generate all products and check them one at a time  
(Sec.~\ref{sec:di_p}), and (2) generate one product simulator that can simulate the behavior of each product of the product line and 
check it in a single verification pass using variability information (Sec.~\ref{sec:di_pl}).

A product line consists of a finite set $\f$ of features and a feature model ${\fm \subseteq \mathcal{P}(\f)}$, 
which defines the set of valid feature combinations (i.e., products).
Each feature $f \in \f$ has an implementation $\impl{f}$ and a specification $\spec{f}$.

\subsection{Detecting Interactions in Products}
\label{sec:di_p}

Once the features of a product $p = \{f_1, \ldots, f_n\} \in \fm$ are selected, 
the composer (e.g., \fh) can generate the corresponding code $\impl{p}$.
The resulting implementation of the product can be checked against the specifications of the features 
selected for the product, that is, $\forall f \in p : \impl{p} \models \spec{f}$.
We use a model checker that \emph{statically} determines whether the execution of the composed product can reach an error state, as defined by the features involved.
If that happens, we know that the composition violates the constraints of at least one participating feature and indicates a feature interaction.

To verify that all products of a product line are free of interactions, 
we have to generate and check all products individually, which we call the \emph{brute-force approach}:

\begin{equation*}\small
\begin{array}{c}
\dfrac{\forall p \in \fm : \forall f \in p : \impl{p} \models \spec{f}}{(\fm, \f)~\mathrm{OK}}
\end{array}
\end{equation*}

\begin{figure}[t]
\centering
\includegraphics[scale=0.65]{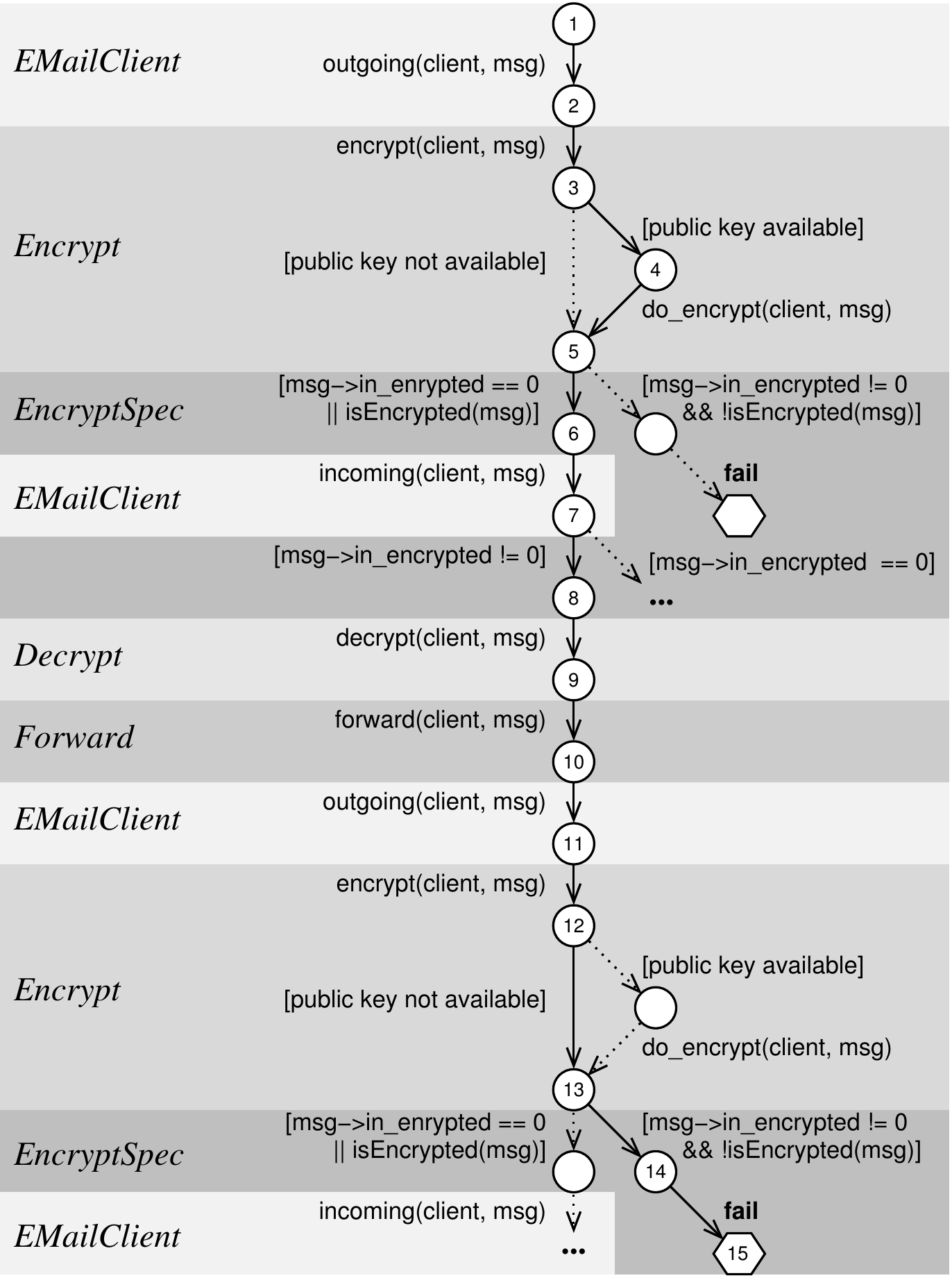}
\caption{Error path of the unsafe interaction between \feature{Encrypt} and \feature{Forward} of the product shown in Figure~\ref{fig:e-mail_client_example}. (\,[...] denotes a condition.)}
\label{fig:art}
\end{figure}

\smallsec{Example}
In Figure~\ref{fig:art}, we show the output of a model checker that has detected the unsafe interaction between the features \feature{Encrypt} and \feature{Forward} in the instrumented code of the product shown in Figure~\ref{fig:e-mail_client_example}.
The figure shows the control-flow graph that describes how the composed e-mail system reaches an error state and thus exhibits an unsafe
feature interaction.
The states along the error path are numbered and connected by solid arrows (dotted arrows do not belong to the error path).
Different parts of the state graph are highlighted with different background colors to illustrate that they belong to different features.
The automaton of feature \feature{Encrypt} (denoted with `\emph{EncryptSpec}' in Figure~\ref{fig:art}) introduces error states (using keyword \code{fail}) that reveal undesired behavior triggered by feature interactions.

\subsection{Detecting Interactions in Product Lines}
\label{sec:di_pl}

An alternative to the brute-force approach is to create a product simulator that contains all feature behaviors of 
a product line, and to use information on their mutual relations during verification. 
The goal is to verify every part of the state space only once, even if the part 
is used in several feature combinations.
The background is that, typically, the products of a product line share many similarities of which a model checker can take advantage, rather than reasoning about individual products in isolation:

\begin{equation*}\small
\begin{array}{c}
\dfrac{\forall f \in \f : \varenc{\fm}{\f} \models \spec{f}}{(\fm, \f)~\mathrm{OK}}
\end{array}
\end{equation*} 

Function $\mathit{var\_enc}$ implements variability encoding, and $\maxp$~denotes the resulting product simulator.
Variability encoding makes information about valid feature combinations from 
the feature-model level available at the implementation level.
All feature code and the feature model are encoded in the product simulator~$\maxp$.
The state space of $\maxp$ subsumes the state spaces of all valid products of the product line,
from which the model-checking procedure can benefit during the verification process.
It allows the model checker to detect feature interactions more efficiently, 
because not all individual feature combinations have to be unfolded 
in the model checker's state space. 

\smallsec{Variability Encoding}
The procedure of variability encoding is a modification of the regular composition process.
All feature modules are composed according to the total composition order.
The resulting product simulator~$\maxp$ can simulate the behavior of
any product of the product line.

First, variability encoding defines for each feature a global boolean variable that models the 
presence or absence of the feature:\,%
\footnote{For simplicity, we assume that there are no name clashes when adding 
the feature variables to~$\maxp$; otherwise fresh names are chosen.}

\begin{equation*}\small
\begin{array}{c}
\dfrac{
f \in \f
}{
\epsilon \longrightarrow \mbox{\sf\footnotesize {\bfseries int}~f};
}
\end{array}
\end{equation*}
where $\epsilon$ is the empty program element.
We model the addition of an element to $\maxp$ as a transformation of the empty element $\epsilon$ to the element
that we want to add.

Second, variability encoding introduces for each function refinement a dispatcher function that dispatches 
between the refined and the refining function depending on whether the feature that 
contains the refinement is selected:
\begin{equation*}\small
\dfrac{
\exists\, m(\overline{P~p'}) \{\, \overline{s'} \,\} \vdash f' \qquad \refines{m}{f}{f'}
}{
\begin{array}{c}
\epsilon \longrightarrow
m(\overline{P~p}) \{\, \mbox{\sf\footnotesize {\bfseries if}\,(}f\mbox{\sf\footnotesize)}~m_f(\overline{p}); \mbox{ \sf\footnotesize {\bfseries else} } m_{f'}(\overline{p});\}\\
m(\overline{P~p'}) \{\, \overline{s'} \,\} \vdash f' \longrightarrow
m_{f'}(\overline{P~p'}) \{\, \overline{s'} \,\}\\
m(\overline{P~p}) \{\, \overline{s} \,\} \vdash f \longrightarrow m_f(\overline{P~p}) \{\, [\mbox{\sf\bfseries\footnotesize original}(\overline{q}) \mapsto m_{f'}(\overline{q})]\,\overline{s} \,\}
\end{array}
}
\end{equation*}
where $m(\overline{P~p}) \{\, \overline{s} \,\}$ is a function declaration with name $m$, 
parameter list $\overline{P~p}$, and
a function body with a list $\overline{s}$ of statements.
The syntax $\mathit{mdecl} \vdash f$ denotes that the function declaration $\mathit{mdecl}$ is introduced (or refined) by feature $f$.
The predicate $\mathit{refines}$ holds if $f$ refines a function~$m$ 
of another feature~$f'$. 
The result of variability encoding is the following:
The refining function is renamed to~$m_f$, the refined function is renamed to~$m_{f'}$,
and the keyword \code{original} is replaced by a call to the refined function.
The dispatcher function uses the value of the boolean variable associated 
with the refining feature $f$ (i.e., whether it is selected or not) to mimic the control flow of a product with 
and without feature $f$ (i.e., calling $m_{f}$ or $m_{f'}$).

Third, variability encoding represents dependencies between features (i.e., the feature model) using 
a boolean formula over the boolean feature variables and encodes corresponding constraints:

\begin{equation*}\small
\begin{array}{c}
\dfrac{
\mathit{formula} 
  = \bigvee_{p \in \fm} \left(\,(\bigwedge_{f \in p} f) ~\land~ (\bigwedge_{f \in \f, f \not\in p} \lnot f)\,\right) 
}{
\epsilon \longrightarrow \mbox{\sf\footnotesize {\bfseries int}~feature\_model() \{\,{\bfseries return} } \mathit{formula} \mbox{\sf\footnotesize \,\}}
}
\end{array}
\end{equation*}

Finally, we enclose the entire program execution in a conditional block that is executed only if the constraints imposed by the feature model are satisfied; this way, execution paths that are associated with invalid feature combinations are not considered by the model checker: 
\begin{equation*}\small
\begin{array}{c}
\dfrac{
}{
\mbox{\sf\footnotesize {\bfseries int}~main() \{\,} \overline{s} \mbox{\sf\footnotesize \,\}} ~\longrightarrow ~ \mbox{\sf\footnotesize {\bfseries int}~main() \{\,{\bfseries if}\,(feature\_model()) \{\,} \overline{s} \mbox{\sf\footnotesize \,\}\,\}}
}
\end{array}
\end{equation*}

\smallsec{Model Checking}
After variability encoding has generated the product simulator~$\maxp$, 
we check $\maxp$ against the specification of all features of the product line.
We initialize the boolean variables of the features using a nondeterministic choice 
such that the model checker must assume that all feature combinations defined by the feature model may occur.
This way, the model checker checks all valid feature combinations (i.e., combinations of feature code)
without generating any individual product.

\begin{figure}[t]
\centering
\begin{lstlisting}[firstnumber=1,frame=top|bottom,basicstyle=\sf\scriptsize,
keywordstyle=\sf\scriptsize\bfseries,commentstyle=\sf\scriptsize\it]{ar}
// one boolean variable per feature
int EMailClient, Forward;

// encoding the feature model
int feature_model() {
  return EMailClient;   // EMailClient && (Forward || !Forward);
}

// dispatch between 'Forward' and '!Forward'
void incoming (struct client *client, struct email *msg) {
  if(Forward) { incoming_Forward (client, msg); } 
  else { incoming_EMailClient (client, msg); }
}

// refinement of method 'incoming' by feature 'Forward'
void incoming_Forward (struct client *client, struct email *msg) { 
	forward(client, msg);
	incoming_EMailClient(client, msg);
}

// base implementation of method 'incoming' by feature 'EMailClient'
void incoming_EMailClient(struct client *client, struct email *msg) { ... }

// base implementation of method 'forward' by feature 'Forward'
void forward (struct client *client, struct email *msg) { ... }

int main(int argc, char **argv) {
  if(feature_model()) { /* start the e-mail client */ }
  return 0;
}
\end{lstlisting}
\caption{Variability encoding of the composition of \feature{EMailClient} and \feature{Forward}.}
\label{fig:e-mail_client_example_lifted}
\end{figure}

\begin{figure}[t]
\centering
\includegraphics[scale=0.75]{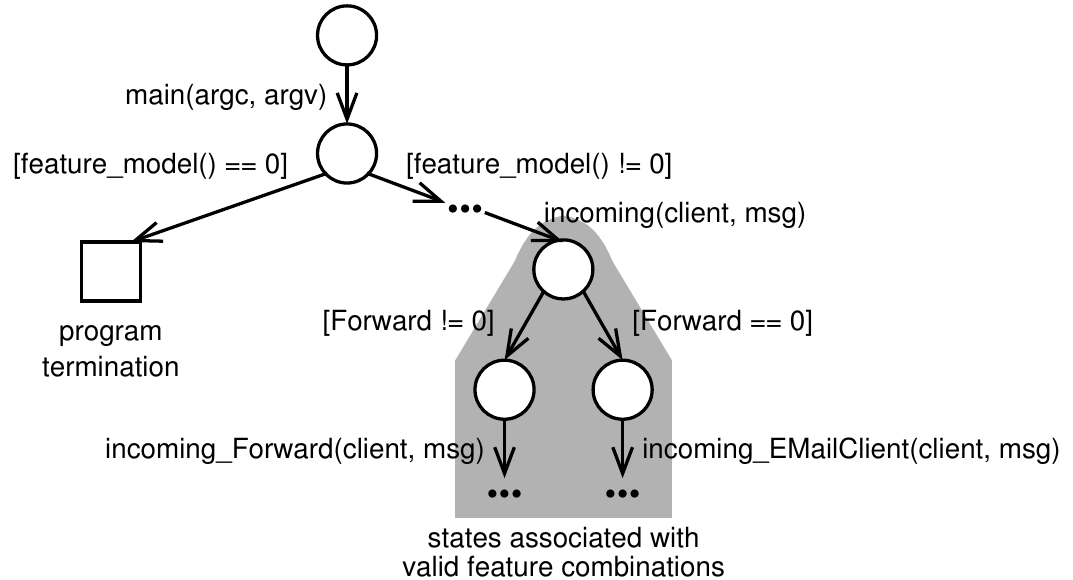}
\caption{State graph of product simulator~$\maxp$ with encoded variability (excerpt).}
\label{fig:lifted_sg}
\end{figure}

\smallsec{Example}
Figure~\ref{fig:e-mail_client_example_lifted} shows the product simulator for the set
\{\feature{EMailClient}, \feature{Forward}\} of features,
as produced by the variability encoding of our tool chain (see Sec.~\ref{sec:impl}). 
Function \code{incoming} (lines~10--13) dispatches between its 
variants with and without feature \feature{Forward}. 
The feature model is encoded (lines~2--7) and the execution is guarded (line~28).

In Figure~\ref{fig:lifted_sg}, we show the effect of variability encoding on the state graph.
States that are associated with invalid feature combinations are not considered by the model checker (left subtree).
All other states are checked. 
Hence, it can be verified that none of the valid feature combinations exhibits an unsafe feature interaction (right subtree). 
Also one can see how both alternative execution paths ---for products with and without feature \feature{Forward}--- 
are encoded in the state graph.

\smallsec{Correctness of Variability Encoding}
In order to use product simulators in feature-aware verification,
we need to show that variability encoding constructs a system that contains precisely
all unsafe feature interactions that are present in any single product (no more, no less).
Therefore, correctness means that a product simulator 
is able to correctly simulate all valid products during model checking. 
 
\vskip 2ex
\noindent\textbf{Theorem} (Correctness of variability encoding):
Given a set~$F$ of features and a feature model~$\fm$,
then the following holds ($\maxp = \mathit{var\_enc}(\fm, F)$):
\begin{equation}\small
\begin{array}{c}
\forall p \in \fm : \forall f \in p :\\ \sel{p}{\maxp} \models spec(f) \iff \impl{p} \models spec(f)
\end{array}
\end{equation}
where $\sel{p}{\maxp}$ configures $\maxp$ to behave like the composition of features~$p$ 
by setting the boolean variables of all features $f\in p$ to $\mathit{true}$ and the rest to $\mathit{false}$.

\vskip 2ex
\noindent\textit{Proof.}
The correctness of variability encoding can only be guaranteed for type-safe product lines (i.e., all products are type correct~\cite{AKG+10}). 
Furthermore, we restrict ourselves to the core operations of feature composition: 
features can add new program elements such as functions, fields, and structures or refine existing functions by overriding.
If there are alternative definitions of a function, field, or structure (due to mutually exclusive features), the respective alternatives must have a common supertype that can be used uniformly in the product simulator --- a property that is also required in certain product-line type systems~\cite{KAT+11}.
As said previously, our composition approach assumes a total composition order, denoted with $\prec$, over the features (cf.~Sec.~\ref{sec:background}). 

We prove that the product simulator simulates all products correctly with respect to the specifications of the features involved, denoted by simulation relation $\sim$\,.
The simulation relation is understood as behavioral equivalence 
after partial-evaluation reduction over the feature variables.
We construct the proof by structural induction on the composition. 
\begin{equation}\small
\label{eq:composition}
\begin{array}{rcl}
\forall v_k \subseteq F_k: \sel{v_k}{\maxp_k} \, \sim \, \bigoplus_{f \in v_k} f
\end{array}
\end{equation}
with $1 \leq k \leq |F| \mbox{ and } f_i \in F \mbox{ and } i < j \iff f_i \prec f_j$\,; 
$\oplus$ denotes regular composition (respecting the global feature order; used inside $\mathit{impl}$);
$F_k$ is a subset of $F$ that contains the first $k$ features with respect to the global feature order (i.e., $F_k = \{f_i \in F~|~i \leq k\}$);
$\maxp_k$ denotes a product simulator consisting of the first $k$ features.

\vskip 2ex
\noindent \textit{Induction base} ($k = 1$):
In the base case, the product simulator consists only of a single feature and 
behaves therefore similarly to the regular product, as no dispatcher functions are introduced:
\begin{equation*}\small
\begin{array}{c}
\sel{\{f_1\}}{f_1}  \, \sim \, f_1
\end{array}
\end{equation*} 

\vskip 2ex
\noindent \textit{Induction hypothesis}:
For all possible configurations of the first $k$ features, the variability-encoded composition of the first $k$ features is equivalent to the corresponding regular composition of the features:
\begin{equation*}\small
\begin{array}{rcl}
\forall v_k \in \mathcal{P}(F_k): \sel{v_k}{\maxp_k} \, \sim \, \bigoplus_{f \in v_k} f
\end{array}
\end{equation*}

\vskip 2ex
\noindent \textit{Induction step}: 
When considering all feature combinations $v_{k+1}$, we can distinguish feature combinations $v_k \cup \{f_{k+1}\}$ in which feature $f_{k+1}$ is selected and combinations $v_k$ in which feature $f_{k+1}$ is not selected.

For combinations in which feature $f_{k+1}$ is not selected, the variability-encoded product contains fields, functions, and structures of feature $f_{k+1}$, but these do not affect the program execution because they are not referenced by other features (type-safety assumption).
If feature $f_{k+1}$ refines a function, the generated dispatcher calls the refined (original) function instead of the refining function, because the feature is not selected:

\begin{equation*}\small
\begin{array}{rcl}
\forall v_{k} \in \mathcal{P}(F_k): \sel{v_{k}}{\maxp_{k+1}} \, \sim \, \bigoplus_{f \in v_{k}} f
\end{array}
\end{equation*}

For combinations in which feature $f_{k+1}$ is selected, the variability encoding as well as the regularly composed product contain the fields, functions, and structures introduced by feature $f_{k+1}$. 
For each function refinement, a dispatcher calls the refining function, which is similar to the function present in the regularly composed product. 
Regardless of feature $f_{k+1}$ being selected or not, the implementation may call the refined function using \code{original}, whose behavioral equivalence follows from the induction hypothesis. 
Therefore, a dispatched function exhibits an equivalent behavior to a regularly composed function, thus:
\begin{equation*}\small
\begin{array}{rcl}
\forall v_{k} \in \mathcal{P}(F_k): \sel{v_{k} \cup \{f_{k+1}\}}{\maxp_{k+1}} \, \sim \, \bigoplus_{f \in v_{k} \cup \{f_{k+1}\}} f
\end{array}
\end{equation*}

As we have investigated all possible combinations of the first $k+1$ features, we have proved Formula~(\ref{eq:composition}) for $1 \leq k \leq |F|$. As $F_{|F|} = \{f_i \in F~|~i \leq |F|\} = F$, we can state:

\begin{equation*}\small
\begin{array}{rcl}
\forall v \in \mathcal{P}(F): \sel{v}{\maxp_{|F|}} \, \sim \, \bigoplus_{f \in v} f
\end{array}
\end{equation*}
which finishes the proof, as the simulation relation $\sim$ implies that, if a specification is satisfied in a simulated product, it is also satisfied in the regularly-composed product and vice versa (i.e., that Equation (1) holds).
\hfill $\square$

\subsection{Discussion}
\label{sec:discussion1}

\smallsec{Separation of Concerns}
Feature-aware verification is based on the idea that features are implemented as separate and composable units, 
and that a feature's specification is local, i.e., it is not aware of all other features of the system.
With our approach, we would like to explore whether and to what extent feature-local specification 
is possible for detecting feature interactions.
Locality is imperative in scenarios without global domain knowledge, such as in distributed feature composition, and it aids program comprehension.

\smallsec{Brute Force vs.\ Variability Encoding}
Both approaches of feature-aware verification have their merits.
The approach of generating individual products and checking them in isolation 
is feasible for a distributed feature composition scenario, 
in which features are developed mostly in isolation and in which global knowledge on valid feature combinations is not available.
It is useful to find unsafe feature interactions quickly in individual products; 
but, for proving the absence of unsafe feature interactions in a product line, all products have to be generated and verified individually.

The technique of checking a product simulator can improve the scalability of 
feature-aware verification if all features are known in advance.
The idea is to encode variability information and dependencies between features into the code base of the product simulator
to make it available to the model checker.
This way, the model checker is able to check a product line once and to guarantee that none of the possible 
feature combinations contains an unsafe feature interaction (according to the specifications). 
Without variability encoding, we would have to generate and check up to $2^n$ products for 
a product line with $n$~features, in the worst case. 
With variability encoding, we have to generate and check only one product simulator that consists of $n$ features.
In our case study, we provide quantitative arguments on when the first or the second approach is superior (see Sec.~\ref{sec:case_study}).

\smallsec{Generality}
Feature-aware verification does not depend on a specific language or tool. 
It allows us to use off-the-shelf model-checking technology, 
rather than expensive and error-prone self-developments or ad-hoc modifications of proprietary model-checking tools. 
In principle, any pair of specification language and model checker can be used, 
and alternative composition mechanisms such as aspect weaving are possible.
The automata language allows us to check safety properties of a system,
which was sufficient in our case study. 
Fairness or liveness properties are currently not supported.

\section{Case Study}
\label{sec:case_study}
To explore the feasibility of feature-aware verification for the detection of unsafe feature interactions, 
we have developed the tool chain \fc and applied it to a case study. 
\fc and the case study are available on the project's Web site.

\subsection{Implementation}
\label{sec:impl}
\fc is based on several existing tools and on tools that we developed for the purpose of feature-aware verification. 
For composition, we use \fh (i.e., we compose features by superimposition).
For model checking, we use the tools CBMC~\cite{CKL04} and CPAchecker~\cite{CPACHECKER}.
Both tools support the verification of safety properties of C code, CBMC by means of bounded, symbolic model checking and CPAchecker either by means of explicit or symbolic model checking.

To implement feature-aware verification, we have developed a translation framework for our automata-based specification language. 
Technically, each specification is rewritten to an ACC aspect.%
\footnote{ACC is an aspect-oriented language extension of C:\\ \url{http://research.msrg.utoronto.ca/ACC/}}
We use the ACC compiler to inject assertions in source-code locations that have been 
specified by the corresponding automaton and that are relevant to a safety property 
(see Figures~\ref{fig:art},~\ref{fig:e-mail_client_example_lifted}, and~\ref{fig:lifted_sg} for examples).
If the model checker finds that an error label is reachable, an unsafe feature interaction is reported.
The automaton along with the error path are passed to the user for debugging, much like in Figure~\ref{fig:art}.

Variability encoding is implemented using \fh's composition facilities. 
The boolean variables for each feature and the function for encoding the feature model are added via superimposition.
The creation of \code{if} guards is realized by modifying \fh's composition rules (e.g., for the composition of function bodies).

\subsection{Case Study: AT\&T E-Mail Client}
A difficulty of finding an appropriate case study is that it has to be complex enough, such that 
realistic feature interactions occur, and not too large, such that we can still trace what happens 
during the detection of interactions.
We decided to base our case study on the e-mail system of Hall~\cite{Hal05},
because it consists of a sufficient number of features, it contains several realistic and unintuitive feature interactions, 
and it has been used before by other researchers in this area~\cite{LKF05}, 
as it incorporates AT\&{}T's domain knowledge on feature interactions in e-mail systems.

The e-mail system consists of 10 features that give rise to 27 feature interactions.
It is divided into a client and a server.
For our case study, we concentrate on the client because, for now, we do not focus on interactions in distributed scenarios, 
which is in line with previous work~\cite{LKF05}.
In Table~\ref{tab:e-mail_features}, we provide information on all features and feature interactions of the e-mail client.
A comprehensive description of the features and their interactions is available in Hall's article~\cite{Hal05}.

\begin{table}
\begin{center}
\addtolength{\tabcolsep}{-0.75ex}

\begin{minipage}{4.9cm}
\begin{tabular}{ll} \toprule
\textbf{Feature} & \textbf{Short description}\\ \midrule
\feature{EMailClient} & basic e-mail client \\ 
\feature{MailQueue} & queuing e-mails \\
\feature{Keys} & key management \\
\feature{Encrypt} & encrypt outgoing e-mails \\
\feature{Decrypt} & decrypt incoming e-mails \\
\feature{Sign} & sign outgoing e-mails \\
\feature{Verify} & verify e-mail signatures \\
\feature{AddressBook} & manage e-mail contacts \\
\feature{AutoRespond} & respond to e-mails \\
\feature{Forward} & forward incoming e-mails \\ \bottomrule
\end{tabular}
\end{minipage}
\hspace{0.1cm}
\begin{minipage}{3cm}
\begin{tabular}{rl} \toprule
\textbf{Id} & \textbf{Feature interaction}\\ \midrule
0 & \feature{Decrypt}, \feature{Forward} \\
1 & \feature{AddressBook}, \feature{Encrypt} \\
3 & \feature{Sign}, \feature{Verify} \\
4 & \feature{Sign}, \feature{Forward} \\
6 & \feature{Encrypt}, \feature{Decrypt} \\
7 & \feature{Encrypt}, \feature{Verify} \\
8 & \feature{Encrypt}, \feature{AutoRespond} \\
9 & \feature{Encrypt}, \feature{Forward} \\
11 & \feature{Decrypt}, \feature{AutoRespond} \\
27 & \feature{Verify}, \feature{Forward} \\ \bottomrule
\end{tabular}
\end{minipage}
\end{center}
\caption{Features and feature interactions of the e-mail client. The interaction ids match the ids of
Hall~\cite{Hal05} (except for id 0).}
\label{tab:e-mail_features}
\end{table}

We implemented the features of the e-mail client in C with \fh~\cite{AKL09fh} following the specification of Hall (including a base program and two helper features).
Furthermore, we included an entry function to trigger events in the client.
Based on the work of Hall, we developed for every relevant feature a specification in the form of one or several automata.
As discussed previously, a key requirement was to specify the features' behavior and safety properties based on local knowledge.

\subsection{Experiments} 
We conducted a number of experiments with the e-mail product line.
First, we generated all of its 40 products and checked them using both CBMC and CPAchecker.
It turned out that with feature-aware verification, we were able to detect all feature interactions of Table~\ref{tab:e-mail_features} based on the feature-local specifications of the input features.
If the model checker does not report a counterexample (i.e., none of the safety properties has been violated), we can be certain that the composition does not contain a feature interaction that violates the specification of the features involved.\footnote{Although CBMC is a bounded model checker, we can use it for proving the absence of interactions in the e-mail client, because it does not contain loops with statically unknown upper bounds.}

Interestingly, we even found an unsafe interaction in \emph{our} implementation that has not been documented by Hall.
It occurs when both features \feature{Decrypt} and \feature{Forward} are selected (id~0 in Table~\ref{tab:e-mail_features}): 
if a host forwards an e-mail automatically to another host that cannot decrypt this e-mail.
This finding encourages us that our approach is useful to detect unknown feature interactions.
Finally, we checked the entire e-mail client product line using variability encoding.
Again, we were able to detect all feature interactions, but without generating all possible feature combinations.

\subsection{Measurements}
\label{sec:measurements}
To further explore their pros and cons, we compare the brute-force approach (i.e., checking all possible products)
with the variability-encoding approach in terms of verification time.
Our case study contains several unsafe feature interactions, so we made the comparison on a per-interaction basis.
Specifically, we measured the runtime needed to find a feature interaction or to report that no feature interaction has been found.
Because every specification is associated with a feature, both approaches need to consider only feature combinations 
that contain this feature; all other combinations trivially cannot violate the specification.

\begin{figure}[t]
\centering
%\vskip -2ex
%\begin{minipage}{5.6cm}
\includegraphics[scale=0.35]{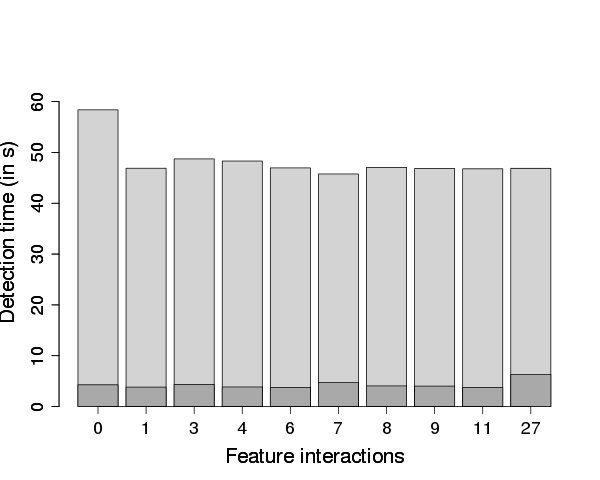}
\caption{Times needed to prove that the individual interactions do not occur 
(brute force in light gray and variability encoding in dark gray).}
\label{fig:safe}
\end{figure}

First, we measured the runtime to prove that a certain interaction does not occur.%
\footnote{We report only results using CBMC in this paper; the entire set of results is available on the project's Web site.}
In Figure~\ref{fig:safe}, we compare the runtime needed to prove the absence of each feature interaction 
for the brute-force approach and the variability-encoding approach. 

%\end{minipage}
%\hspace{0.25cm}
%\begin{minipage}{5.6cm}
%
\begin{figure}[t]
\centering
\includegraphics[scale=0.35]{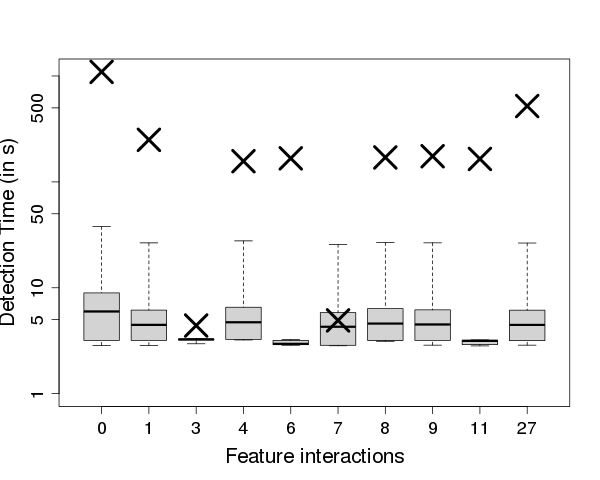}
\caption{Times needed to find the individual interactions 
(brute force as box plots and variability encoding as crosses; y-axis in log scale).}
\label{fig:results}
\end{figure}
%\end{minipage}
%\hspace{0.25cm}
%\begin{minipage}{5.6cm}
%
\begin{figure}[t]
\centering
\includegraphics[scale=0.35]{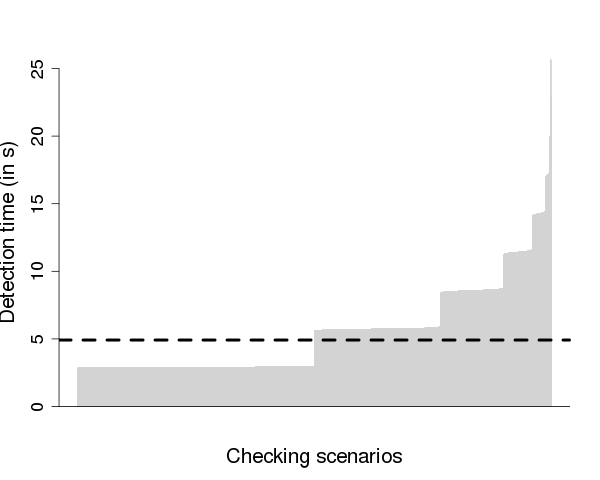}
\caption{Times needed to find the unsafe feature interaction between \feature{Encrypt} and \feature{Verify} 
(brute force as bars and variability encoding as dashed line).}
\label{fig:one_res}
%\end{minipage}
\end{figure}

Second, we measured the runtime to discover each feature interaction. 
For variability encoding, we measured the runtime to find an unsafe feature interaction 
by checking the product simulator against a specification that is violated.
For the brute-force approach, we need to generate and check all possible products in a predefined order.
After the first erroneous feature combination has been identified, no further checks are necessary.
The absolute runtime to find an interaction depends on the order in which the products are checked.
It may be that, incidentally, we choose an order that exhibits an unsafe interaction early 
(e.g., the product that we check first contains the unsafe interaction), such that we obtain the result rather quick.
Or, it may be that only the very last product that we check contains the unsafe interaction, 
such that the runtime to detect the interaction is the sum of the runtimes for checking all possible products of the product line.
In order to remove the bias of the ordering, we perform the following calculation:
Let $\pi = \langle p_1, ..., p_i, ..., p_n\rangle$ be a sequence of $n$~products being checked
such that product~$p_i$ is the first product that violates the specification.
We first measure the runtime $t(p_j)$ to check each possible product~$p_j$ for the considered interaction.
The actual total runtime of sequence~$\pi$ is the sum $c = \sum_{j=1}^i t(p_j)$ of runtimes.
This way, we calculate the total runtime $c_k$ for every possible permutation of sequence~$\pi$
(i.e., the checking order), and obtain the values $c_1, ..., c_q$ of total runtime until finding the unsafe interaction
(given that $q$ is the number of permutations).%
\footnote{Since the number $q$ of permutations is a huge number,
we compute an approximated result, by classifying the runtime values $t(p_j)$
into classes of similar runtime.
This dramatically reduces the number of permutations to consider.}

In Figure~\ref{fig:results}, we show for each unsafe interaction (x-axis) a cross that denotes 
the runtime (y-axis) needed to detect the considered unsafe interaction using the product simulator,
and a box plot that contains all possible total runtimes $c_1, \ldots, c_q$ needed to detect the interaction with the 
brute-force approach.%
\footnote{The box contains 50\,\% of the values.
the two thin lines are the maximal and minimal values; the thick line is the median.}
For illustration, Figure~\ref{fig:one_res} shows a bar plot of the runtimes $c_1, \ldots, c_q$ 
needed to find the interaction between \feature{Encrypt} and \feature{Verify} and 
compare it to the runtime of five seconds that is needed using the product simulator, displayed as a horizontal dashed line.

\subsection{Interpretation}
The first observation is that variability encoding outperforms the brute-force approach by a factor of ten 
in proving the absence of feature interactions (Fig.~\ref{fig:safe}).
In all experiments, the measured time includes only the verification runtime, not the runtime for generating the product
or product simulator, as it is negligible.
The reason for the superiority of variability encoding is that 
every reachable state has to be visited in order to establish a correctness proof.
A model checker has more potential for optimization on the product simulator, 
because all information is available in the system,
compared to the brute-force approach, where the verification process is restarted from scratch for every single product.
For example, if there are similarities between individual products 
(which is a goal of product-line engineering), 
a model checker does not need to check the similar parts repeatedly.

A second observation is that for detecting an unsafe interaction in a faulty product line,
the brute-force approach is substantially faster (factor of 10 to 100) compared to 
checking the product simulator (Fig.~\ref{fig:results}).
This result was not to be expected, especially taking previous results of product-line model checking into account (see Sec.~\ref{sec:related_work}). 
After a careful analysis, we identified several factors that decide on the appropriateness of variability encoding.
First, the product simulator contains the code of all features, 
the feature model, and the alternative execution paths of all possible feature combinations.
Hence, it is more complex than any of the products without variability encoding, 
which increases the complexity (and thus the runtime) of model checking immediately.
However, in the brute-force approach, the challenge is to determine a proper order of checks to minimize checking runtime.
Second, the ratio between the number $b$ of products that contain the interaction and the number $n$ of all possible products influences the benefit of variability encoding.
For bigger values of $b/n$ (close to 1) it is more likely that we pick an order that requires only a few checks to find the interaction.
For example, the interaction between the features \feature{Encrypt} and \feature{Decrypt} (6) occurs in all 40 products.
Hence, the probability $p$ that an interaction is found in the first product is $40/40 = 1$: at most one product needs to be checked to detect the interaction.
In such a situation, we cannot take advantage of variability encoding, as illustrated in Figure~\ref{fig:results}.
But for lower values of $b/n$, it is less likely that we choose a product checking order that exhibits the unsafe interaction.
For example, with a ratio of $8/40$, the probability to find the unsafe interaction between \feature{Encrypt} and \feature{Verify} with the first check is $1/5$.
In such a situation, we can benefit from variability encoding.

\subsection{Lessons Learned}
\label{sec:ll}
The lessons we learned by applying feature-aware verification to the e-mail client product line
can be summarized as follows.

We were able to detect all documented feature interactions of the e-mail client based on C code and automata-based specifications.
We even found a previously undocumented interaction, which encourages us that our approach is able to detect unknown interactions in other applications.

Although feature interactions occur between two or more features, we were able to detect them based on feature-local specifications.
That is, there is no global knowledge necessary to detect them.
Locality is not only imperative for scalability or distributed feature composition, but it is a prerequisite to detect undocumented feature interactions.

Due to the potentially large number of feature combinations in product lines, interaction detection based on model checking is more expensive than in monolithic systems.
We have demonstrated that variability encoding can reduce the verification runtime by a factor of ten in proving the absence of feature interactions in the e-mail client. 
The reason for this reduction is the reuse of partial verification results
when checking the product simulator.

The ratio between the number of feature combinations that contain an interaction and the number of all possible feature combinations influences the benefit of variability encoding. 
With lower values ---which are to be expected in practice--- it is superior to the brute-force approach. 
A careful combination of the brute-force approach and the variability-encoding approach seems to be favorable. 
The results suggest that generating and checking some products is useful in early verification stages to discover unsafe interactions quickly, and that the importance of variability encoding increases in later development stages when the number of interactions decreases.
Compared to previous work, this insight is a significant step toward understanding and improving product-line model checking.

\subsection{Study Limitations}
We conducted a case study to explore whether feature-aware verification can be used for feature-interaction detection in a non-trivial and controlled setting. 
Although a rigorous empirical study is currently elusive, we still gained a number of interesting insights (see Sec.~\ref{sec:ll}), which shall encourage us and others to follow this line of research. 

\section{Related Work}
\label{sec:related_work}

The feature-interaction problem was explored for different domains
in the literature~\cite{CKM+03}.
Our work addresses the problem in the context of feature orientation and product lines. 
There are two approaches in the literature that address the combinatorial explosion of feature combinations
in software product lines: 
(1) check features as far as possible in isolation and (2) check the entire product line in a single pass.

The first approach has been explored by Li et al.~\cite{HKF02,LKF05} and Liu et al.~\cite{LBL11}.
They propose to verify features modularly based on formal transition systems and CTL.
The idea is to check as much as possible at the level of individual features to save effort when checking their compositions.
Verifying a feature, it is determined which parts of a specification the feature satisfies and which parts have to be satisfied by other features. 
This information constitutes a semantic interface of the feature, which is used during the verification of its composition with other features.
Li et al.\ and Liu et al.\ have a slightly different verification scenario in mind: they check to what extent a feature satisfies a specification that a product has to fulfill.
In our approach, each feature comes with its own specification that states which properties have to hold when the feature is selected.

The second approach (i.e., check an entire product line in a single pass) has been explored by Lauenroth et al.~\cite{LTP09} and Classen et al.~\cite{CHS+10,CHS+11}.
Lauenroth et al.\ have developed an extended model checking approach that takes product-line variability into account~\cite{LTP09}.
Similarly to variability encoding, their approach is able to verify that every valid product that can be derived from the product line fulfills certain properties.
In contrast to variability encoding, they require to extend the model-checking tool to incorporate variability information.
They evaluated their work by means of two examples based on I/O automata and CTL, not on the basis of program code.

Classen et al.\ have developed a model-checking technique for the verification of
feature-extended transition systems against temporal properties~\cite{CHS+10}.
In principle, their approach is similar to the approach of Lauenroth et al.; 
it is based on an extension of the model-checking algorithm.
They have developed a model-checking tool in Haskell and applied it to check a mine-pump controller consisting of nine features.
They report substantial performance gains over individual product verification, 
but did not recognize the influence of the ratio between products that contain an unsafe interaction and all products of a product line. 
In a recent extension of their approach, Classen et al.\ use a system modeling language with explicit feature support and encode information on features into the transition system~\cite{CHS+11}. 
However, they do not encode the feature model, and they do not support the verification 
of software written in a mainstream programming language.

Based on the work of Lauenroth et al.\ and Classen et al., we explored whether and how product-line verification techniques can be used for feature-interaction detection. 
We specially considered the implementation and specification of features in separate and composable units, 
which was not the focus of Lauenroth et al.\ and Classen et al. 
Finally, we pursue an approach that is based on off-the-shelf model checking, 
rather than on self-developments and extension of existing model checkers.

Post and Sinz proposed the notion of configuration lifting to verify variable C code efficiently~\cite{PS08}.
The background is that it is usually too expensive to generate all possible configurations of a C file that contains preprocessor directives such as \code{\#ifdef}. 
The idea is to replace each conditional preprocessor directive by a corresponding \code{if} statement 
thus making it accessible to a software verification tool. 
Variability encoding is based on their approach.
Although we use it in a different scenario (feature composition instead of conditional compilation), 
the main aspects are similar.
With regard to the work of Post and Sinz, we contribute formal arguments and a proof of the correctness of variability encoding, 
which is especially important in a feature-composition scenario.

\section{Conclusion}
Feature-aware verification is an approach to detect unsafe feature interactions in feature-oriented product lines.
We implement \emph{and} specify features in separate and composable units, and detect unsafe interactions based on feature-local specifications.
Locality of feature specifications is important for scalability and distributed feature composition.
We used, extended, and developed a tool chain that supports feature-aware verification based on off-the-shelf model checking technology, and we presented a formal model (along with an argument for the correctness) of variability encoding.
We were able to automatically detect critical feature interactions (including a previously undocumented interaction) 
in Hall's e-mail system~\cite{Hal05}.

Variability encoding aims at improving the verification performance for software product lines.
Rather than generating and checking all possible feature combinations, 
we encode variability and dependency information into the product simulator's code base and check it in a single pass.
In our e-mail case study, variability encoding saved up to 90\,\% of the checking time in proving the absence of interactions, 
but is slower than the brute-force approach if many products contain an unsafe interaction.
An insight, compared to previous work, is that 
variability encoding is superior if the task is to verify the absence of unsafe feature interactions.

\section*{Acknowledgments}
%%%db: Alphabetically.
We are grateful to J.~Atlee, A.~Classen, and M.~Rosenthal for their comments to earlier drafts of this paper. 
We thank S.~Boxleitner for his C implementation of the e-mail client.
This research was supported in part by the German DFG grants AP~206/2 and AP~206/4,
and by the Canadian NSERC grant RGPIN 341819-07.

\balance

\end{document}